\documentclass[9pt,twocolumn,twoside]{opticajnl}
\journal{opticajournal} 

\setboolean{shortarticle}{true}

\usepackage{lineno}
\usepackage{textcomp}
\usepackage{parskip}
\usepackage{textgreek}
\usepackage{xcolor}

\newcommand \red[1]{#1}

\title{Wavelength tuning of \emph{VCSELs} via controlled strain}

\author[1,2,*]{Salah Guessoum}
\author[1,2]{Athanasios Kyriazis}
\author[1]{Tushar Malica}
\author[1]{Jürgen Van Erps}
\author[2]{Geert Van Steenberge}
\author[1]{Martin Virte}

\affil[1]{Brussels Photonics Team (B-PHOT), Vrije Universiteit Brussel, Pleinlaan 2, 1050 Brussels, Belgium}
\affil[2]{Center for Microsystems Technology (CMST), Ghent University and imec, Technologiepark-Zwijnaarde 126, 9052 Ghent, Belgium}

\affil[*]{salah.eddine.guessoum@vub.be}

\begin{abstract}
Besides major advantages for telecommunication applications, VCSELs have attracted interest for their potential for neuro-inspired computing, frequency comb generation or high-frequency spin oscillations. In the meantime, strain applied to the laser structure has been shown to have a significant impact on the laser emission properties such as the polarization dynamics or birefringence. 
In this work, we further explore the influence of strain on VCSELs and how this effect could be used to fine tune the laser wavelength. Through a comprehensive investigation, we demonstrate consistent wavelength shift up to 1 nm and report a sensitivity between 0.12 to 0.18 nm / millistrain. We also record birefringence values up to 292 GHz. Our results show that controlled strain level could be considered for fine wavelength tuning and possibly alleviate the selection of VCSEL for precise wavelength requirements.
\end{abstract}

\setboolean{displaycopyright}{false} 

\begin{document}

\maketitle


Vertical-Cavity Surface-Emitting Lasers play a pivotal role in the field of optoelectronics, particularly in applications like short-reach optical communication and sensing \cite{Liu:19,Iga_2008}. Their cost efficiency and low power consumption make them indispensable components. It has been proven that it is possible to modify the output optical properties of VCSELs through thermal strain \cite{Pusch:Thermal_Birefringnce,Choquette:94} and through anisotropic strain along specific crystallographic axes \cite{Krassi:in_place_anisotropes,VanDerSande:2006_stress_temp_effect_VCSELs}. This approach allows the selective modification of the lattice constants of the semiconductor material through the elasto-optic effect \cite{Chen:elatstooptic}. These changes enable the manipulation of optical properties such as polarization switching \cite{Ostermann:2007}, polarization chaos \cite{TRaddoPolChaos:2017} and wavelength tuning as a consequence of the change in the optical susceptibility \cite{VanExter_Biref_VCSEL}. Inspired by in-plane anisotropic strain, more recent engineering techniques explore controlled strain application using microactuators found in piezo-electric materials like lead zirconate titanate (PZT) or nano and microelectromechanical systems (MEMS) \cite{Huang:07, Shi:20}. These approaches require modifications in the VCSEL structure. It is crucial to follow the correct integration procedure and exercise precise control over the micro-actuator systems. \red{While achieving significant wavelength tuning capabilities of 2.5nm for GaAs-based VCSELs \cite{Huang:07} and 6nm for 930nm VCSELs with electroplated copper bases \cite{Shi:20},these methods often involve complex fabrication procedures. Thermal strain wavelength tuning was also reported at 35 pm/K \cite{Nakahama_2014} and 68 pm/K \cite{Ti_Xue_HP_temperature_VCSEL_tune_2022:}.}\\
In our study, we propose a novel approach to wavelength tuning of VCSELs, thereby, bypassing the need for changing the intricate fabrication processes of the VCSELs. Focusing on the impact of controlled mechanical strain on VCSEL wavelength characteristics, our investigation utilizes a custom four-point bending module to explore strain-induced wavelength tuning. This encompasses a detailed examination of wavelength evolution and polarization of the output light. Our measurements were conducted on two arrays of 1x4 VCSELs coming from the same batch, and we demonstrate the repeatability of results across the different VCSEL chips of both arrays. Understanding the interplay between mechanical strain and VCSEL performance not only deepens our knowledge of the behavior of these lasers but also unlocks new opportunities for their application in high-speed optical communication systems, where strain-induced wavelength and birefringence changes \cite{M.LindemannStrain_Spin_VCSEL:2019} can play a crucial role in enhancing modulation bandwidths and transmission speeds.

\begin{figure}[t!]
\centering
\centering\includegraphics[width=\linewidth]{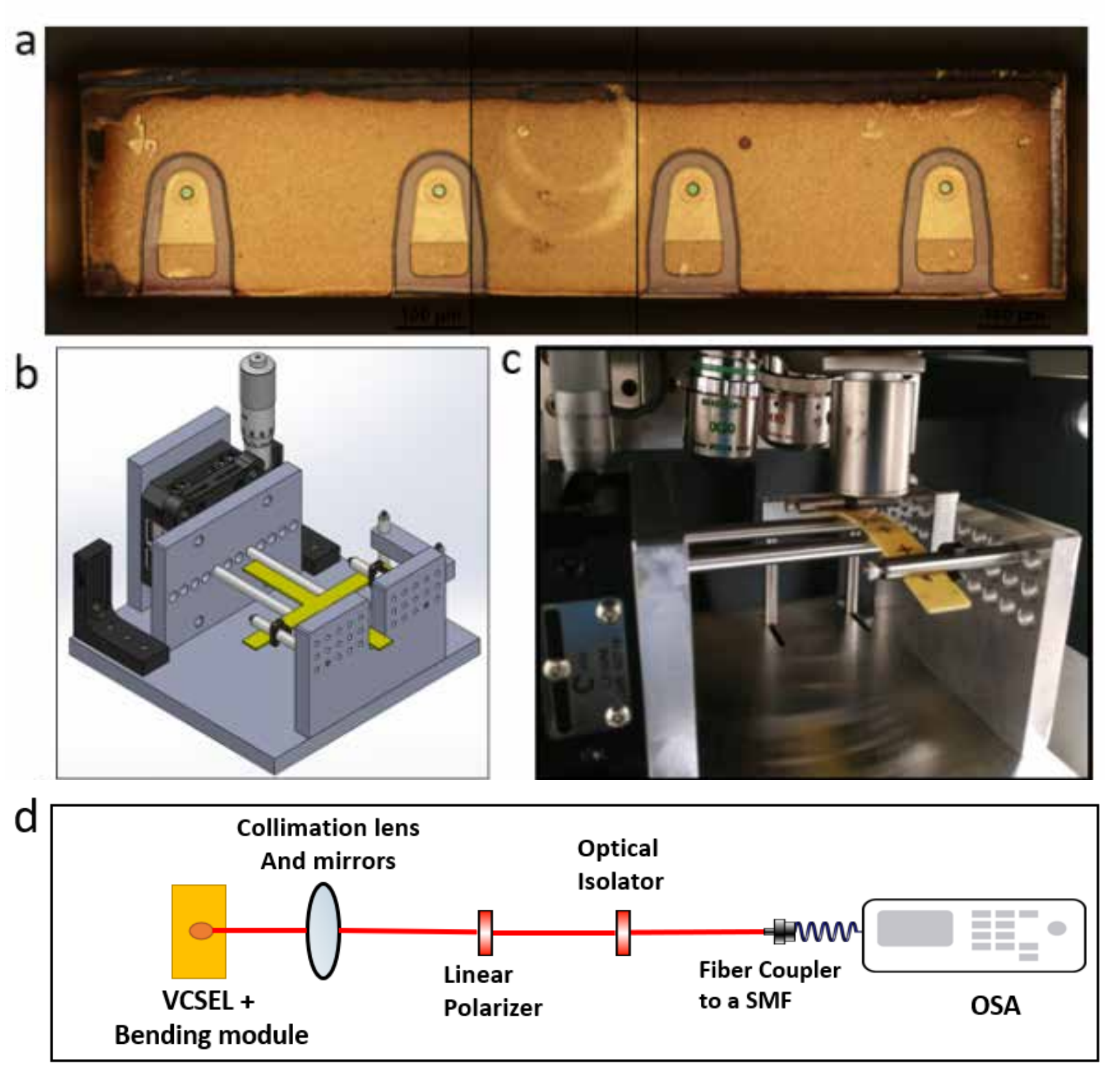}
\caption{Experimental setupa) 1x4 VCSEL chip array integrated on top of the substrate. The circular marks visible on the center of the sample are attributed to a vaccuum pickup tool employed to place the VCSEL array ontop of the PCB substrate. b) Schematic of the 4-point bending module used to apply strain onto the VCSELs. c) A strained VCSEL sample held under a white light interferometer. d) simplified schematic of the optical setup used to measure the optical properties of a strained VCSEL.}
\label{fig:ExperimentalSetup}
\end{figure}

For our experiments, we used two samples that we name Sample-A and Sample-B, each containing a 1x4 array of 1550 nm VCSELs grown on the same wafer, as shown in Fig.\ref{fig:ExperimentalSetup}(a). The lasers are described in details in \cite{Muller:11}. We refer to the VCSELs on the samples as VCSELs- A1-4 and B1-4. These VCSELs are fixed on top of a bendable substrate made of FR-4 material which is also a printed circuit board material. To ensure the VCSEL array is centered on the substrate, a patterning was drawn on the substrate to better position the array. The strain is applied to the substrate and transferred onto the VCSEL chips. To ensure strain transfer from the substrate to the VCSEL chips, we attached the VCSEL array using a thermal-cured adhesive at 150°C (EPO-TEK 353ND). This adhesive ensures a comprehensive bonding of the VCSEL array to the substrate which allows strain transfer. As for the connections, on Sample A, copper fanouts wire bonded to the VCSELs have been added. On Sample B, however, we probed the VCSELs directly. As detailed later in the text, such a difference did not have any significant impact on the qualitative results of our measurements.
The experiments were performed in a thermally controlled environment with an ambient temperature of 20\textdegree C.
Any strain experienced by the VCSELs during the integration process such as of thermal or mechanical origin is considered as the reference or the resting state of the system.

The strain is applied to the VCSELs via a custom four-point bending module, see Fig.\ref{fig:ExperimentalSetup}(b) and (c). This module consists of a solid base for system stability and four rods representing the loading pins of the bending technique. The rods are supported by two plates. One plate is stationary, while the other is connected to a translation stage with a micrometer screw. \\
In the 4-point bending method, the sample is positioned horizontally across two supports (in our case, the stationary rods), with two additional rods (connected to the translation stage) exerting load onto the sample through controlled translation. This setup generates a bending moment inducing a deformation of the sample. To correctly apply the bending moment, the sample should be well-centered and the rods placed symmetrically compared to the center of the sample. The advantage of using the four-point bending technique is to ensure that the sample between the two inner loading pins is subjected to a constant bending moment. The bending moment is translated to strain and stress on the VCSELs.This technique ensures that the 4 VCSELs within the array are subjected to similar strain levels. \\
The module we use is robust and ensures reproducibility of the applied strain, the position of the substrate ensures a vertical emission of the VCSELs and the dimensions of the module allow the possibility of using WLI to estimate the strain, the estimation will be further explained in the next paragraphs.

Fig.\ref{fig:ExperimentalSetup}(d) shows the experimental setup used in our study. It employs the aforementioned VCSEL samples placed on our a custom-made 4-point bending module (See Fig.\ref{fig:ExperimentalSetup}(b)). A 3mm focal length lens collimates and aligns the light beam. The light is focused onto a single-mode fiber (SMF) connected to a high-resolution optical spectrum analyzer (APEX Technologies AP2083A), and measurements were acquired at a spectral resolution of 0.8 pm.
A linear polarizer is also employed to control the polarization of the beam and to verify strain-induced polarization dynamics. An optical isolator prevents optical feedback from the surface of the SMF.\\
The experiment involves varying the strain applied on the two samples and measuring the laser output of the VCSELs for different current values. During the straining of the samples, the translation movement of the bending module leads to a change in the position of the VCSELs. In order to correctly accommodate this change, the collimation lens was placed on a translation stage. Next, probe needles were placed and removed before and after each measurement to prevent mechanical damage to the connection pads. Optical spectra of the different VCSELs were collected for varying current values at each strain level. The measurement spectra were also used for power estimation.\\
To prevent damage to our samples during the measurements, we limited the screw displacement to a maximum value of 4mm. Once the limit was reached, we reset the micrometer screw to its initial position to experiment on a new VCSEL. The initial position, which represents the state where no strain is applied to the sample, is referred to as the "rest state", experimentally the position of the screw corresponding to the rest state is the position for which we start feeling counter force while rotating the actuator; this force is due to the resistance of the substrate to the mechanical stress applied.\\
To estimate the strain, an independent measurement was conducted in which the sample and bending module were placed under a white light interferometer (WLI)(see fig\ref{fig:ExperimentalSetup}(c)). Using the data of the WLI we are able to compute the radius of curvature (R) of the PCB substrate. Given the low thickness of the VCSEL array compared to the substrate, since we consider a total strain transfer between the substrate and VCSEL array, the strain is linked to the radius of curvature by \textepsilon = y / R where \textit{y} is defined as the distance from the neutral axis to the surface of the substrate. We consider \textit{y} to be half the thickness of our substrate, i.e. 0.5mm. A linear fit of the data collected gives an estimate of the strain evolution as a function of the screw displacement at about 1.9 millistrain/mm with a resolution of 0.19 \textmu Strain.

\begin{figure}
	\centering
	\centering\includegraphics[width=\linewidth]{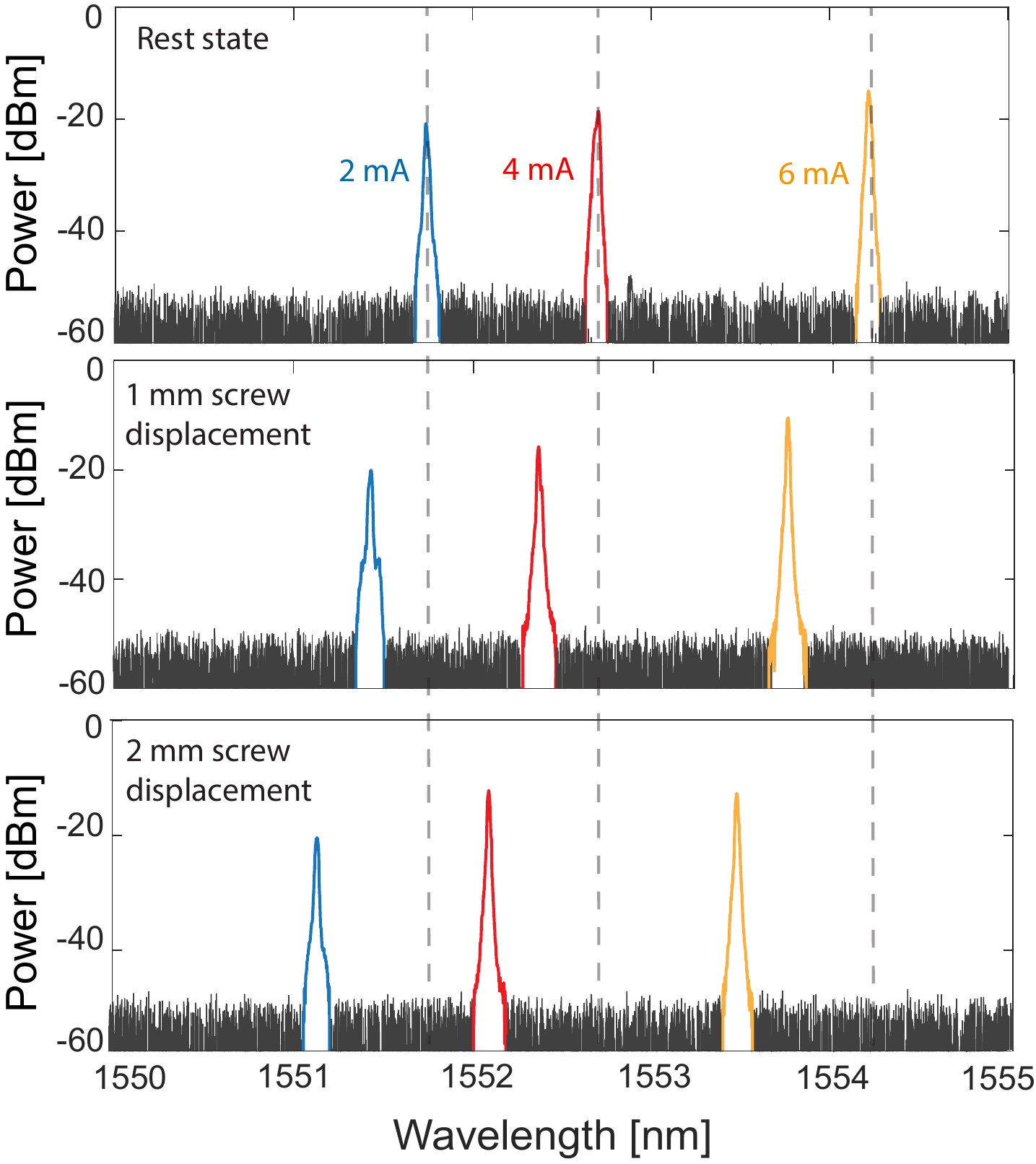}
	\caption{Spectra of VCSEL A1 measured as power (dBm) at different current values (mA) under progressive strain levels. The perpendicular dotted lines mark the peak positions of the rest state revealing the strain-induced blue shift of the VCSEL wavelength.}
	\label{fig:Spectra_frames}
\end{figure}

\begin{figure}
	\centering
	\centering\includegraphics[width=\linewidth]{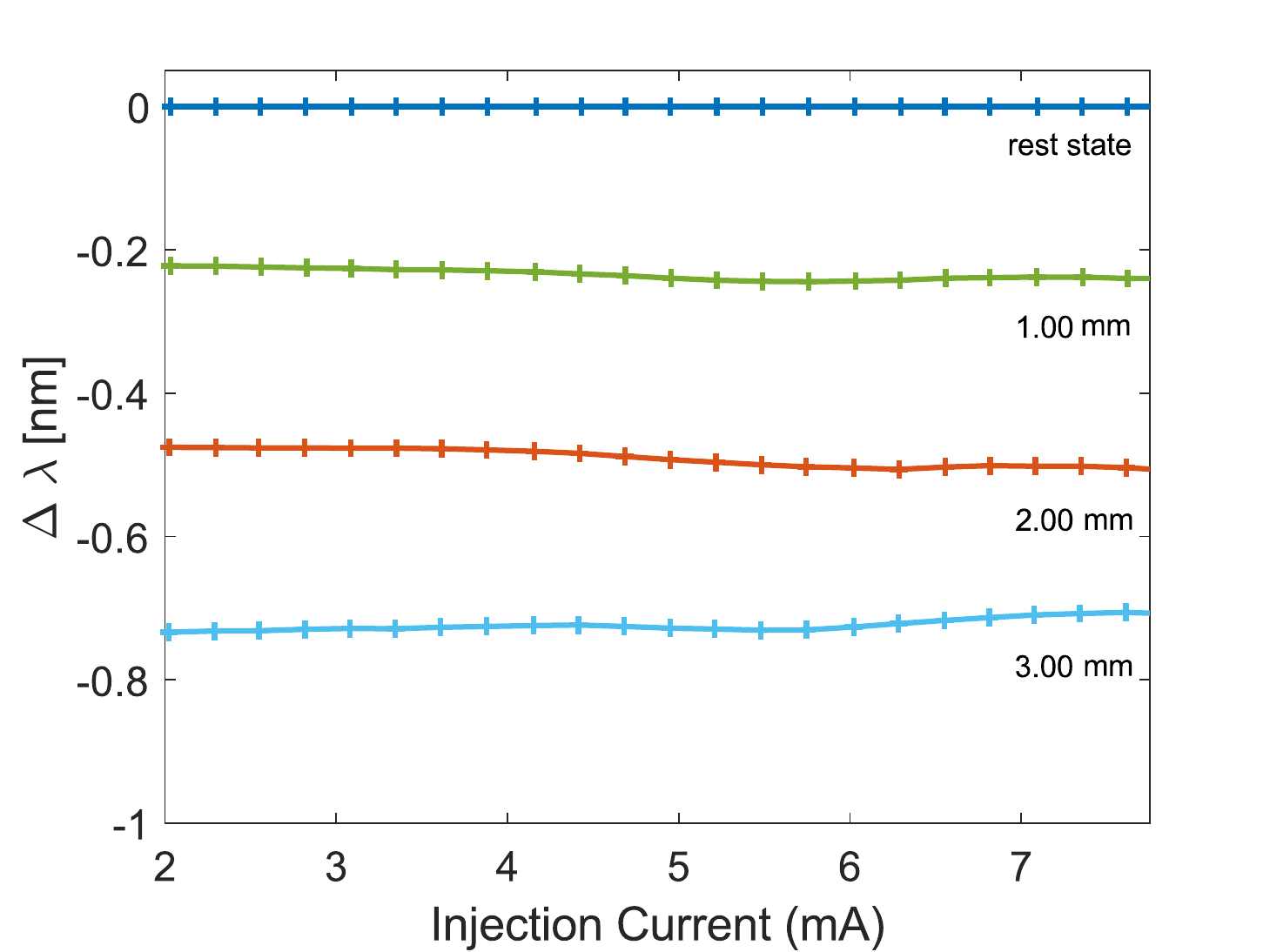}
	\caption{Evolution of wavelength shift (nm) with respect to the rest state as a function of the injection current (mA) for a range of strain levels.}
	\label{fig:Horizontal_lines}
\end{figure}

Here, we present the results of our measurement showing the effect of strain on the VCSEL emission. as mentioned in the description of the experimental setup, our data is extracted from the spectra recorded using the Apex OSA.
No significant power drop or change in the behavior of the optical output of the VCSELs at different strain levels was observed. For each strain level, the average variance in power is 0.38 dBm which we mainly attribute to slight differences in the optical alignment from one measurement to another.\\
Next, we focus on characterizing the wavelength change of the VCSELs under different strain levels.
In Fig.\ref{fig:Spectra_frames}, the recorded spectra of VCSEL A1 for current values from 2 mA to 6 mA, and under different levels of strain, are shown. The frames in Fig.\ref{fig:Spectra_frames} correspond to three different states: rest state, 1 mm of screw displacement, and 2 mm of screw displacement respectively. By comparing the frames, we notice a blue shift in the measured wavelength of the peak indicative of strain-induced changes. This observed blue shift is present among all our test VCSELs. Moreover, the relative wavelength change is similar between the VCSELs.\\
Fig. 3 shows the difference between the measured peak wavelength for each strain level with respect to the reference state, where we consider the rest state spectra as the reference. The observed blue shift appears to be unaffected by current variations, also suggesting that device temperature might have only a limited impact on this behavior. This leads us to believe that the blue shift is induced mainly by the applied strain. We notice a progressive wavelength shift as the screw displacement was increased, indicating a direct link between the strain and the induced wavelength shift. We attribute the small fluctuations observed in normalized wavelength shifts reported in Fig.\ref{fig:Horizontal_lines} to the minor fluctuations in the temperature of the environmental conditions. The aforementioned observations and strain-induced dependency were observed to be consistent across all samples. The results from Fig.\ref{fig:Spectra_frames} and Fig.\ref{fig:Horizontal_lines} imply a reproducible, controllable, and effective strain-induced wavelength tuning technique while maintaining a relatively stable output power.

\begin{figure}
	\centering
	\centering\includegraphics[width=\linewidth]{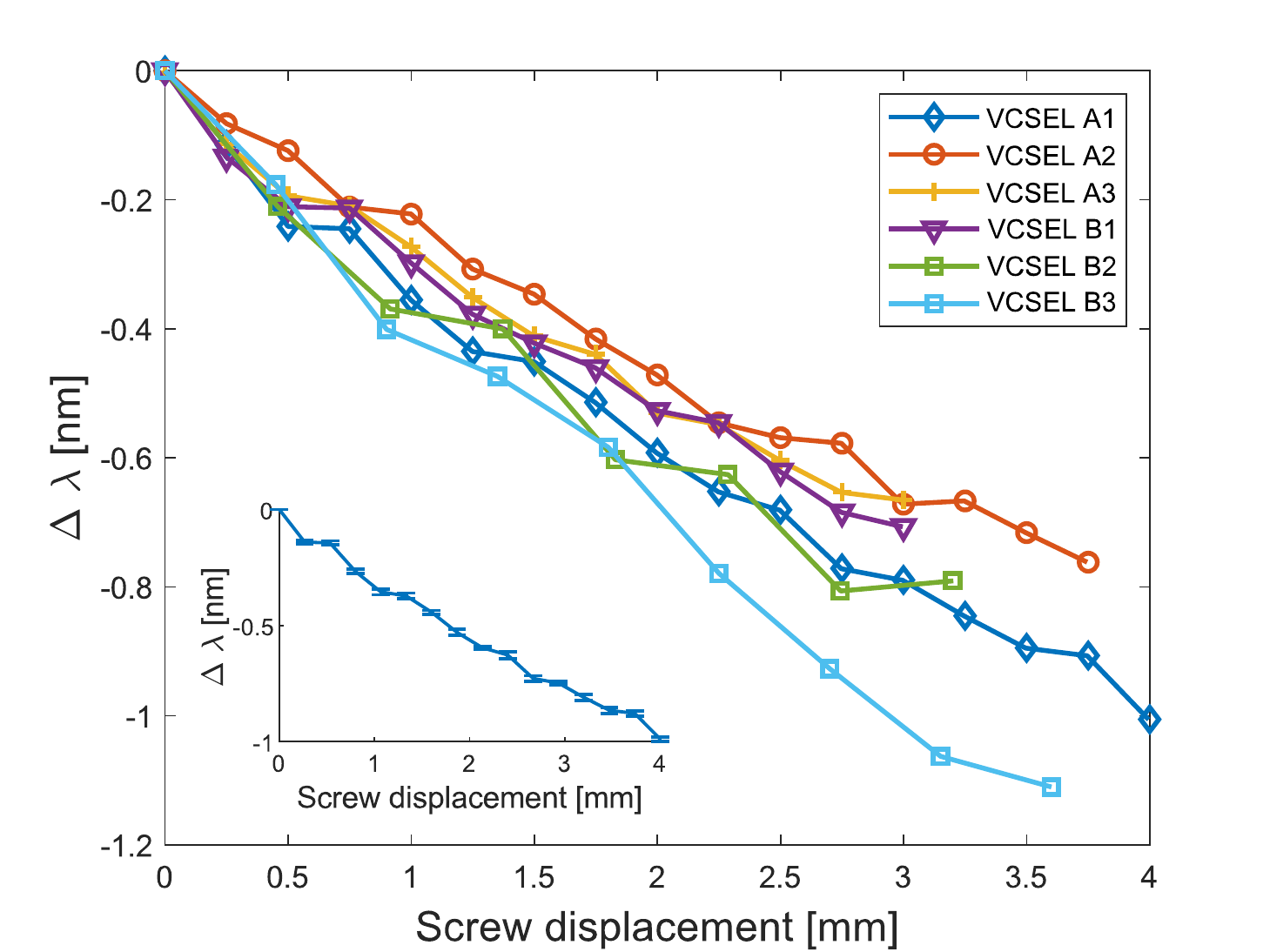}
	\caption{The average wavelength shift (nm) of each VCSEL experiment across all injection currents (mA), plotted as a function of screw displacement (mm). The inset: a magnified view of the wavelength shift (nm) for the VCSEL A1, \red{with error bars included, on average around +/- 11 pm, showing low variance.}}
	\label{fig:All_WL_shift}
\end{figure}

Fig.\ref{fig:All_WL_shift} shows the evolution of the measured wavelength shift of all sample VCSELs as a function of the level of screw displacement. The data presented in the plot is obtained by averaging all driving current measurements between 1 mA and 8 mA of our  VCSELs for the different screw displacement levels, from the rest state to 4mm of displacement. The variance of the data over the driving current values is negligibly small as can be in seen the inset figure of Fig.\ref{fig:All_WL_shift} which represents the data of VCSEL A1 with the calculated error bars. The average variance for VCSEL A1 is reported at 11 pm. The error bars were intentionally excluded from the main plots for the sake of visibility.
The evolution of the wavelength of each VCSEL in the figure follows a linear dependency with screw displacement. By computing linear fits on top of the collected data we obtain an approximation of the linear dependency in nanometers/millimeter of screw displacement as shown in Table.\ref{tab:VCSEL_WL_shift}. \red{With a displacement step of 0.2 mm, we didn't observe hysteresis when decreasing or increasing the deformation. The deformation appears to be reversible and reproducible.} The consistent linear dependency of our measurements among the different VCSELs suggests the potential of reliable wavelength tuning through controlled mechanical strain, further strengthening the use of this approach in optimizing VCSEL performance for applications in telecommunication \cite{Iga_2008,Liu:19} demanding precise wavelength tuning and a consistent power output.

The strain-induced blue shift can be attributed to the elasto-optic effect as well as to the direct deformation of the laser cavity \cite{Chen:elatstooptic,Krassi:in_place_anisotropes,Ostermann:2007}. The elasto-optic effect refers to the change in the refractive index of a material in response to applied mechanical stress. The changes in the refractive indices along the crystallographic axes influence the birefringence within the VCSEL and therefore an observable wavelength change \cite{Pusch:Birefringence_Splitting,VanExter_Biref_VCSEL}. The structure of the VCSELs used in our experiments contains a Mesa structure on the aperture \cite{Muller:11} that stabilizes the polarization of the VCSEL emission by further suppressing the suppressed mode of the VCSEL \cite{Spiga:Mesa2017}. In our experiments we could only observe the dominant and suppressed linear polarized modes for one of our VCSELs, namely VCSEL A3. By analyzing the evolution of the wavelength change with strain, we see that the wavelength of both modes changed in different directions, a blue shift for the dominant mode as shown in Fig.\ref{fig:All_WL_shift} and a red shift for the suppressed mode, the wavelength change means having an increased splitting between the two modes. We estimate the evolution of the frequency splitting between the two modes to be 100GHz for the rest state up to 292GHz for a screw displacement of 3mm. Our measurements align with the birefringence estimations reported in \cite{Pusch:Birefringence_Splitting,M.LindemannStrain_Spin_VCSEL:2019}.\\
The direct deformation of the laser cavity can lead to changes in birefringence as well as a modification of the band gap energy of the semiconductor material. Direct deformation can therefore induce wavelength shifts\cite{Shi:20}, enhance carrier confinement, and improve optical gain\cite{Huang:07} which we do not explicitly observe during our measurements. 
\red{We also acknowledge that prolonged exposure to strain could yield fatigue effects, potentially changing the response of the VCSEL to the applied strain. Even though we did not observe such effect during our study, we only subjected each VCSEL devices to a dozen strain cycles at most and therefore we cannot draw conclusions on the impact of fatigue at this stage.}

\begin{table}[htbp]
\footnotesize 
\caption{\bf Summary of results indicating the 6 VCSEL wavelengths \textlambda \space (nm) and estimation of the linear evolution of wavelength for applied strain (nm /mm), the third column uses our calculated approximation of strain levels to determine the wavelength evolution per applied strain (nm/ mStrain)}
\begin{tabular}{|c|p{2cm}|p{2cm}|p{2cm}|}
\hline
VCSEL  & \textlambda \space [{nm}] & \textDelta \textlambda \space per screw displacement [{nm}/{mm}] & \textDelta \textlambda \space per applied strain [{nm}/{mStrain}] \\
\hline
A1 & 1551.7 & 0.288 & 0.1515\\
A2 & 1548.5 & 0.226 & 0.119\\
A3 & 1550.5 & 0.243 & 0.128\\
B1 & 1549.4 & 0.252 & 0.1326\\
B2 & 1550.4 & 0.282 & 0.1484\\
B3 & 1550.7 & 0.34  & 0.179\\
\hline
\end{tabular}
  \label{tab:VCSEL_WL_shift}
\end{table}

Our study introduces a new method to apply controlled mechanical strain to VCSELs in a systematic way, that is by integrating the VCSELs on top of a bendable substrate and by employing a four-point bending module we are able to finely control the strain level we apply to the VCSEL. We have demonstrated that mechanical strain can shift the VCSEL's wavelength by up to 1 nm consistently. This wavelength shift occurs progressively across increasing levels of strain, indicating the direct and reliable impact of mechanical strain on the VCSEL's optical characteristics. Our findings provide valuable insights into the impact of mechanical strain on VCSEL behavior. By clarifying the relationship between applied strain, resulting birefringence changes, and emitted wavelength, we offer valuable guidance for optimizing VCSEL performance in practical applications.
The ability to manipulate the emission wavelength precisely through direct manipulation of strain holds great promise for various optical communication and sensing applications. By leveraging the effects of strain on VCSELs, researchers and engineers can enhance device performance and functionality, leading to advancements in optoelectronic systems.
In conclusion, our study highlights the importance of considering mechanical strain as a direct and effective means of manipulating the VCSEL's optical characteristics. This work opens up new avenues for exploring the potential of strain-induced effects in VCSEL technology, paving the way for continued innovation in the field.

\begin{backmatter}

\bmsection{Acknowledgments} 
The authors acknowledge the support of the Research Foundation - Flanders (FWO, Grant number G020621N ), the European Research Council (ERC, Starting Grant COLOR’UP 948129, MV) and the METHUSALEM program of the Flemish government.

\bmsection{Disclosures} 
The authors declare no conflicts of interest.

\bmsection{Data Availability Statement} 
The data are available from the corresponding author upon reasonable request.

\end{backmatter}

\bibliography{sample}

\begin{thebibliography}{10}
\newcommand{\enquote}[1]{``#1''}

\bibitem{Liu:19}
A.~Liu, P.~Wolf, J.~A. Lott, and D.~Bimberg, \enquote{Vertical-cavity surface-emitting lasers for data communication and sensing,} {\protect\JournalTitle{Photon. Res.}} \textbf{7}, 121--136 (2019).

\bibitem{Iga_2008}
K.~Iga, \enquote{Vertical-cavity surface-emitting laser: Its conception and evolution,} {\protect\JournalTitle{Japanese Journal of Applied Physics}} \textbf{47}, 1 (2008).

\bibitem{Pusch:Thermal_Birefringnce}
T.~Pusch, E.~La~Tona, M.~Lindemann, \emph{et~al.}, \enquote{{Monolithic vertical-cavity surface-emitting laser with thermally tunable birefringence},} {\protect\JournalTitle{Applied Physics Letters}} \textbf{110}, 151106 (2017).

\bibitem{Choquette:94}
K.~D. Choquette, D.~A. Richie, and R.~E. Leibenguth, \enquote{{Temperature dependence of gain‐guided vertical‐cavity surface emitting laser polarization},} {\protect\JournalTitle{Applied Physics Letters}} \textbf{64}, 2062--2064 (1994).

\bibitem{Krassi:in_place_anisotropes}
K.~Panajotov, B.~Nagler, G.~Verschaffelt, \emph{et~al.}, \enquote{{Impact of in-plane anisotropic strain on the polarization behavior of vertical-cavity surface-emitting lasers},} {\protect\JournalTitle{Applied Physics Letters}} \textbf{77}, 1590--1592 (2000).

\bibitem{VanDerSande:2006_stress_temp_effect_VCSELs}
G.~Van~der Sande, M.~Peeters, I.~Veretennicoff, \emph{et~al.}, \enquote{The effects of stress, temperature, and spin flips on polarization switching in vertical-cavity surface-emitting lasers,} {\protect\JournalTitle{IEEE Journal of Quantum Electronics}} \textbf{42}, 898--906 (2006).

\bibitem{Chen:elatstooptic}
C.~L. Chen, \emph{Appendix D: Elasticity, Photoelasticity, and Electrooptic Effects} (John Wiley and Sons, Ltd, 2006), pp. 421--436.

\bibitem{Ostermann:2007}
J.~M. Ostermann, P.~Debernardi, A.~Kroner, and R.~Michalzik, \enquote{Polarization-controlled surface grating vcsels under externally induced anisotropic strain,} {\protect\JournalTitle{IEEE Photonics Technology Letters}} \textbf{19}, 1301--1303 (2007).

\bibitem{TRaddoPolChaos:2017}
T.~Raddo, K.~Panajotov, B.~Borges, and M.~Virte, \enquote{Strain induced polarization chaos in a solitary vcsel,} {\protect\JournalTitle{Scientific Reports}}  (2017).

\bibitem{VanExter_Biref_VCSEL}
M.~P. van Exter, A.~K. Jansen~van Doorn, and J.~P. Woerdman, \enquote{Electro-optic effect and birefringence in semiconductor vertical-cavity lasers,} {\protect\JournalTitle{Phys. Rev. A}} \textbf{56}, 845--853 (1997).

\bibitem{Huang:07}
M.~C. Huang, Y.~Zhou, and C.~J. Chang-Hasnain, \enquote{Nano electro-mechanical optoelectronic tunable vcsel,} {\protect\JournalTitle{Opt. Express}} \textbf{15}, 1222--1227 (2007).

\bibitem{Shi:20}
J.-W. Shi, Z.~Khan, R.-H. Horng, \emph{et~al.}, \enquote{High-power and single-mode vcsel arrays with single-polarized outputs by using package-induced tensile strain,} {\protect\JournalTitle{Opt. Lett.}} \textbf{45}, 4839--4842 (2020).

\bibitem{Nakahama_2014}
M.~Nakahama, H.~Sano, S.~Inoue, \emph{et~al.}, \enquote{Wavelength tuning and controlled temperature dependence in vertical-cavity surface-emitting lasers with a thermally and electrostatically actuated cantilever structure,} {\protect\JournalTitle{Japanese Journal of Applied Physics}} \textbf{53}, 010303 (2013).

\bibitem{Ti_Xue_HP_temperature_VCSEL_tune_2022:}
X.~Ii, Y.~Zhou, X.~Zhang, \emph{et~al.}, \enquote{High-power single-mode 894 nm vcsels operating at high temperature (> 2 mw @ 365 k),} {\protect\JournalTitle{Applied Physics B}} \textbf{128} (2022).

\bibitem{M.LindemannStrain_Spin_VCSEL:2019}
M.~Lindemann, X.~Gaofeng, T.~Pusch, \emph{et~al.}, \enquote{Ultrafast spin-lasers,} {\protect\JournalTitle{Nature}} \textbf{568}, 212--215 (2019).

\bibitem{Muller:11}
M.~Muller, W.~Hofmann, T.~Grundl, \emph{et~al.}, \enquote{1550-nm high-speed short-cavity vcsels,} {\protect\JournalTitle{IEEE Journal of Selected Topics in Quantum Electronics}} \textbf{17}, 1158--1166 (2011).

\bibitem{Pusch:Birefringence_Splitting}
T.~Pusch, M.~Lindemann, N.~Gerhardt, \emph{et~al.}, \enquote{Vertical-cavity surface-emitting lasers with birefringence splitting above 250ghz,} {\protect\JournalTitle{Electronics Letters}} \textbf{51}, 1600--1602 (2015).

\bibitem{Spiga:Mesa2017}
S.~Spiga, D.~Schoke, A.~Andrejew, \emph{et~al.}, \enquote{Effect of cavity length, strain, and mesa capacitance on 1.5-μm vcsels performance,} {\protect\JournalTitle{Journal of Lightwave Technology}} \textbf{35}, 3130--3141 (2017).

\end{thebibliography}

\bibliographyfullrefs{sample}

\end{document}